# Rationality-proof consensus: extended abstract

Jean-Philippe Martin (self) and Eunjin (EJ) Jung (RationalMind)

# Abstract

Blockchain systems benefit from lessons in prior art such as fault tolerance, distributed systems, peer-to-peer systems, and game theory. In this paper we argue that blockchain algorithms should tolerate both *rational* (self-interested) users and *Byzantine* (malicious) ones, rather than assuming all non-Byzantine users are *altruistic* and follow the protocols blindly. Such algorithms are called *BAR-tolerant* [1]. To design a BAR-tolerant system, one can follow these three steps: clearly define the utility function for the rational users, prove the algorithm is such that there is no benefit from unilaterally deviating (that is, it's a Byzantine Nash Equilibrium), then prove the algorithm correct assuming the rational actors follow the protocol. We present an example attack by rational users: the *gatekeeping* attack, where members of a system selfishly decide to prevent newcomers from joining. This attack may affect any stake-based system where the existing members prevent newcomers from making a stake, and essentially form a cartel. We then sketch a BAR-tolerant consensus protocol for blockchain that can defend against this attack. It relies on a strict order to decide who gets to propose a new block (so there's no need to race to solve a crypto puzzle) and it relies on hardware ID tokens to make sure every computer is only represented at most once as a block proposer to mitigate Sybil attacks. It also defends against the gatekeeper attack. The BAR-tolerant approach is naturally also applicable to other blockchain algorithms.

# Introduction

Blockchain algorithms have captured the world's imagination, and are considered for applications beyond payment services. As they grow, we want to make sure we design these algorithms carefully so they can work reliably even at a large scale. Reliability and scalability are not new challenges in computer science so naturally we draw from prior art including distributed systems, peer-to-peer, and game theory. The latter especially in the context of incentive compatibility, a key property for Bitcoin or similar systems where software isn't run by a trusted central authority: the participants might write a modified version of the software, one that deviates from the specified algorithm in order to gain some unfair benefit to the user. This selfish behavior is happening already: some Bitcoin miners skip the validation of transactions even though this could result in including invalid transactions in the ledger. Research shows that consensus in Bitcoin will be difficult to achieve when the block reward becomes so small that the transaction fees are the main source of incoming for miners [6].

BAR tolerance [1] is the property of protocols that work despite both selfish users (called *rational*) and some number of malicious users (called *Byzantine*). We argue that BAR-tolerance is a desirable property for blockchain systems. In this paper, we discuss an example attack on the system by rational users and show how a BAR-tolerant blockchain protocol might defend against it. We sketch this protocol to show how one would go about designing a BAR-tolerant algorithm; our protocol is based on the consensus from [1].

A *gatekeeping* attack is an example attack by rational users in blockchain systems that requires a stake (security deposit). More precisely, this attack is applicable to any system that requires a user to make a deposit before participating in any protocol that gives rewards proportional to the stake. Many Proof-of-Stake consensus algorithms, including those used in Ethereum Casper FFG and Tendermint, use deposit-slashing [5] as a solution to the Nothing-at-Stake problem [4], and require a stake from a user before participating in consensus. For a new user to become a block producer and earn the block rewards, the new user has to deposit a stake first to a special address, which is a transaction in the chain's native currency. For the existing block producers, the expected block reward they could earn decreases as the newcomer joins. If a block producer is chosen with a probability proportional to its own stake over the total stake in the system, then as a new comer joins, the total stake increases and the probability of being elected decreases. If the block producers vote on candidates and get the reward proportional to its own stake over the total stake, then as a new comer joins, the total stake increases and the amount of reward for each block decreases. Thus it is a rational behavior for the existing block producers to not include the deposit transaction in their blocks, which prevents a newcomer from becoming a block producer.

## Related Work

Blockchain systems experience attacks from Byzantine users. Bitcoin Gold suffered from double spending attack [9]: The hacker with the majority of the computing power was able to deposit the same Bitcoin Gold to an exchange and also to its own wallet. The attacker may have stolen $18.6 million from the exchange. Monacoin also experienced a selfish-mining attack that cost an exchange $90K [7]. Software bug could also cause a Byzantine error in the system: overflow in sum created more bitcoins for block reward than what the protocol dictates and also two addresses received 92.2 billion bitcoins each [3].

While it is less evident than Byzantine attacks, the community has been speculating that the rational users are the majority of the system and may deviate from a desirable behavior. For example, some bitcoin miners generated invalid blocks [2]: they would rather spend the time to check the validity of transactions on solving the hash puzzle. Researchers in [6] showed that it may be rational to fork the chain when there are not many transactions left, and the transaction fees form the majority of the reward.

# Approach

We need to tolerate both malicious and self-interested (rational) actors.

Since some actors may be malicious, we need to use a Byzantine-tolerant algorithm. These algorithms can function despite arbitrary behavior from some bounded fraction of the participants (the exact fraction depends on the model. It's typically 1/3, but can be up to 100% if sufficiently strong assumptions can be made). The key requirement is to bound the fraction of malicious actors (otherwise they may be able to evict honest actors, prevent correct transactions from entering the blockchain, etc.). Instead of using a proof of work (that is, relying on the assumption that the aggregate computing power bought by the honest actors is larger than that of the malicious actors), one can use modern hardware features such as the TPM chip present in most motherboards today. The manufacturer guarantees actors cannot pretend to own more motherboards than they genuinely do. So long as the aggregate buying power of the honest actors is larger than that of malicious actors, the majority vote of the system will be in honest hands - without needing to continually consume vast amounts of power. These hardware tokens prevent Sybil attacks. Remains only to pick who can add the next block to the blockchain: this could be done at random, or simply by taking turns.

Since some actors can be rational (perhaps everyone who is not malicious), we need an incentive-compatible algorithm. To tolerate both Byzantine and rational actors we use the BAR approach [1]. The idea is to design a protocol that is a Byzantine Nash Equilibrium. This means that it is in the best interest of each rational actor to follow the protocol as specified. In this methodology one first needs to specify what the rational actors will consider *costs* and *benefits* when computing their utility function. A protocol is a Byzantine Nash Equilibrium if rational actors see no increase in their utility from unilaterally deviating from the protocol. We assume that they consider the Byzantine actor's worst possible behavior when estimating the utility they would get from following a particular sequence of actions. To prove a BAR-tolerant protocol correct, we show
   1. That the algorithm is a Byzantine Nash Equilibrium, and then
   2. That the algorithm has the desired properties, under the assumption that the rational actors obey the protocol.

The BAR model allows (but does not require) some of the actors to be altruistic, meaning that they follow the protocol even if it weren't in their personal best interest to do so. In this paper we do not require (nor take advantage of) altruistic actors. The model also assumes that rational actors do not collude. This doesn't mean that no collusion happens, it just means that if some actors collude then they count against the limit on Byzantine actors. When designing a BAR-tolerant protocol, the author chooses what counts as costs or benefits. This is a trade-off: including more things as costs or benefits means it applies to more individuals in real life, but it also makes for a craftier rational actor and the protocol needs to be correspondingly stronger (and its correctness argument gets correspondingly longer). Including only a few things (for example, perhaps the rational actor is only trying to optimize how many tokens they earn) keeps the algorithm and proof simpler, but some individuals may not fit this model. This isn't as bad as it sounds: those actors that act rationally but have different incentives than those the system was designed to resist will still be tolerated, they will just count against the Byzantine tolerance threshold.

We believe that BAR-tolerance or similar approaches are a good fit for blockchain algorithms because the high stakes make the temptation to write a modified client harder to resist (so rational behavior may be seen in practice), and the same stakes also mean that it's important to rely on as few assumptions as possible (so tolerating arbitrary behavior from some actors is beneficial).

# Key benefits

We sketch a blockchain algorithm that can (a) tolerate malicious actors without requiring a wasteful proof of work, and (b) tolerate rational actors, acting in their individual self-interest. We hope it serves as an illustration of how one may design a BAR-tolerant algorithm in the context of a blockchain.

Proofs of work in Bitcoin waste large amounts of electricity, creating unnecessary pollution. They defend against *Sybil attacks*, where a single bad actor takes on multiple identities in order to have more voting power and subvert the protocol. Proof of stake has been proposed as an alternative; our approach is similar but with small stakes, allowing everyone to join and be equally rewarded for participating in the chain. Hardware features (e.g. TPM) can be used to hinder bad actors trying to pretend to be more than one person: they have to buy real hardware for each voter so it's not as easy as just sending another network message.

Another function of Bitcoin's proof of work is to act as a lottery, to pick a leader. Here instead we use a round-robin approach, taking turn. We rely on a BAR-tolerant protocol to ensure that we can reach consensus on which block to add to the chain, without giving rational actors room to deviate for gain.

# Components

The protocol is built modularly, out of multiple component parts. Going in approximate chronological order, a user can first read and write transactions in the chain. Then if they want to produce blocks (and be rewarded for it), they go through a joining and registration step. Then they follow the consensus subprotocol to wait for their turn and finally produce a block.

Keeping the protocol modular helps keep the design simple and the proofs manageable, though a few of the arguments still need to be over the system as a whole. It also makes it easier to evolve the system over time and to add features. For example in this paper we are not touching on smart contracts or other high-level concepts, but of course they can also benefit from being Byzantine and rational tolerant.

## Joining and Registration

Any user of the system can submit transactions to the blockchain without having to register or join, but they need to do both in order to produce blocks. Joining is done by submitting a transaction that transfers a small amount of "coin" to a special address for members (this address may be managed by a smart contract, or be hardcoded into the consensus protocol itself). The user can get this deposit back when they're done producing blocks (this may be done automatically when the user closes the mining program). The deposit is forfeit if the user is found deviating from the protocol.

The registration protocol is part of the same interaction, but with the purpose of making it harder for a single malicious person to create multiple accounts in an attempt to manipulate voting or other aspects of the system, i.e. mitigating Sybil attack. Registration can be done by reading a unique hardware ID from the user's machine, so that each user is only allowed a single identity in the system. For example one may use the key pair embedded in a TPM, or SGX's linkable quotes to ensure that a given computer cannot register more than once. Another approach is to ask the user to pay a one-time fee of some sort. The fee could be a computation (compute some hashes), human endeavor (mail in a postcard), or paying a token amount. Something that is small enough that users don't mind doing it once, but that would become a burden if it had to be done thousands of times.

One of the subtleties of the joining component is that we want it to work even if an existing block producer benefits from preventing a particular user from joining. This works out with a bit of care as each block producer in turn gets the power to admit the new user. In our model, a rational user assumes that all other rational users follow the protocol, so it expects other rational block producers would let this particular user join. In other words, as block producers take turns admitting new users, a rational block producer can delay a particular user from joining when it is its own turn to admit new users, but cannot prevent it indefinitely. Another subtlety is the question of whether adding a block producer reduces everyone's reward, and so everyone may be incentivized to delay new users joining the block producers group. This is a rational behavior not only for our consensus protocol but also for any Proof-of-Stake consensus algorithm where the staking (join) operation needs to be included in a blockchain, such as Ethereum Casper FFG as of May 2018 or Tendermint. One way to prevent delayed joins is to make sure everyone's reward remains the same as others join. Since with more participants there is a longer wait between turns, that means we'd have to keep on increasing the individual reward, which will lead to inflation. Instead, the protocol can detect attempts at delaying a join, and punish that behavior. The rational choice is then to accept the joiner since the punishment is worse than the benefit from delaying the joiner.

We incentivize the block producers to include a join operation by both rewards and punishments. The reward is that the join transactions do not count toward the block size limit, so the block producers can include as many join transactions as they wish to maximize their gain from the transaction fees. The punishment happens when a block producer does not include the join transaction in a new block it is proposing even though the transaction was sent to it in advance. Participants who wish to join must send their request to multiple actors, and have them forward the request to a proposer. Since joins do not count against the block size limit, this proposer is always able to include the join transaction. If they do not, each of the relay actors will sign a SUSPECT message and send it to the next leader. This next leader will strip the faulty proposer of their block reward if they have recent SUSPECT messages from f+1 distinct actors (so Byzantine actors by themselves are not able to force punishment). A participant might choose to send their join request to only a single proposer at a time instead of multiple, choosing to accumulate SUSPECT messages over time instead. Either way, it is in the block proposer's best interest to publish the join transaction since the cost of the punishment is larger than the benefit from delaying that particular user from joining.

## Block Production

Block producers have the job of taking in transaction requests from users, putting them into a block, and linking it to the blockchain. When things go well, the producers simply take turns. When things don't (for example in case of failure or asynchrony), the BAR-tolerant consensus protocol ensures that a new producer is picked and that every non-malicious actor still agrees on a unique block to append to the chain.

The other facet of block production is making sure that all user transactions are included. We can use a similar mechanism as for join: producers can be suspected of misbehavior and punished if they ignore a user for too long. This ensures that rational users are disincentivized from trying to delay a user's transaction.

## Shared State

In order for the block producers to take turns, as described in the previous section, we need them to agree on who the block producers are. For this, we use blockchain order of the "join" and "leave" transactions. Both transactions (paying to join or withdrawing deposits to leave) are recorded in the blockchain. Since the blockchain implements an append-only log, everyone agrees on the sequence of membership transactions (joins and leaves) so they agree on who the block producers are and in which order they were added. Thus, everyone agrees on a queue of block producers: they take turns producing a block, so for each block everyone knows who the next block producer should be.

Producers can be kicked out if they misbehave and this is also done as a transaction in the blockchain so we can leverage the same mechanism to keep track. Rational block producers have an incentive to keep track of this state since it's necessary to follow the protocol correctly and avoid punishment or expulsion (or just missing their turn).

It may be beneficial to share more state, such as the balance of every user if the blockchain is used to keep track of a cryptocurrency balance. A similar mechanism can be used, but other variants are possible as well such as having a different group of actors keep track of the state, or offering a "summary" service that allows one to quickly catch up to the present state of the system.

## Consensus

The consensus protocol ensures that a unique block is added to the chain. More formally:

Each participant ("block producer") may propose a value ("block"). At the conclusion, each participant that follows the protocol ("correct participant") will decide on a block, and
- the block was proposed by a participant, and
- all correct participants decide the same block.

In a long enough period of synchrony, the protocol will always conclude. In periods of asynchrony it may or may not conclude. If it does, the guarantees hold.

It uses the shared state described above to determine which block producer is the "leader", in charge of publishing a block. If this leader is too slow then it switches to the next leader. The algorithm ensures that even in periods of asynchrony, no two different values are decided.

We run an instance of consensus for each block to add to the chain, keeping track of the current leader so everyone gets their turn. Blocks are linked together in the traditional way, with a hash, but each block also includes signed messages from the protocol participants that vouch for its correctness. This prevents a malicious participant from forging a block.

As an additional consideration, instead of taking turns in a predictable order (1, 2, 3, ...) we can use a VRF-like function [8] to pick the next leader based on the contents of the last block. This still gives everyone the same number of turns on average, but it makes it difficult for an adversary to run a denial of service attack against the next leader since its identity is only revealed at the last minute.

## Details about consensus

To make things concrete, let's flesh out the consensus part. At a high level, block producers take turns and a timeout mechanism ensures eventual progress. We start by recapitulating our assumptions.

## Model

We assume partial synchrony: the system has asynchronous periods where there is no bound on message delivery or processing time, and it also has synchronous periods with a bound on both. We pick a timeout T that is long enough for consensus to finish in synchronous periods. To formalize the notion that "synchronous periods are long enough to make progress" we prove that the algorithm works if the system is eventually forever synchronous. In addition, we assume clock rates are similar, so that at the scale of a single consensus instance, if two users have clocks t1 and t2, their ratio is bounded by a constant r: 1/r <= t1/t2 <= r.

We assume that if messages are sent infinitely often, they are eventually delivered. Using acknowledgements and resends this can implement reliable links between actors that follow the protocol.

We have Byzantine and rational actors, as per the BAR model. We make no assumption about the behavior of malicious actors. For the rational actors we assume that they are not forced to obey the protocol as it was presented to them but instead they can follow any different algorithm of their choosing. However, they only do so if it increases their net utility from participating in the system. This utility function accounts for the costs and benefits from using the system (e.g. coins earned and spent, but also the ability to make transactions at all). For this protocol we are only considering the following for the utility function: coin balance, the ability to exclude specific transactions, the ability to force their own transactions to be included in the blockchain, and the ability to include transactions that would normally be rejected (such as double-spending). Naturally, while the rational actor would see double-spending their own money as a benefit, the protocol guarantees this cannot happen.

We assume there are at most one third of malicious actors. The rest are rational. We use "f" for the number of malicious actors and "n" for the total number of block proposers, so n>3f. We are not assuming that any of the actors are altruistic (meaning they follow the protocol blindly), although of course if some are the system still works.

## Structure

The protocol follows the classical three-phase commit approach and has "rounds". In round R, actor (R modulo n) is the leader. In each round, either the leader sends its value to everyone (who then eventually decide it), or no value is decided. If no value is decided then the next round starts. This round first "freezes" the previous round, preventing it from making further progress. The freeze process also determines whether a block might have been decided in the previous round. If so, the new leader must propose that block and finish the job. This new leader will be able to propose their own block in the instance of consensus that follows immediately.

In a round, the communication pattern goes like this:

- Leader sends an AGREE message to everyone. This message is signed and includes the proposed block. The leader then collects signed acknowledgments for the hash of the block.
- Leader sends a WRITE message to everyone. This signed message includes 2f+1 of the previous acknowledgments for the same hash. Since users ack only one message, this guarantees that the leader cannot send a block to some users and a different one to some other users. The users store the WRITE message and acknowledge.
- Leader sends a DECIDE message to everyone. This signed message includes 2f+1 signed acks from the previous round. Users decide when they see this message.

If the current leader is too slow (perhaps it's Byzantine, or has crashed), we start the next round. This happens after f+1 actors suspect the current leader of being too slow and send a signed TIMEOUT message to the next leader. To ensure this next round R doesn't conflict with the previous one, we "freeze" round R-1 before doing so. The freezing algorithm is a single exchange:
- New leader sends FREEZE to everyone. This signed message includes the f+1 TIMEOUT messages. The users that have not yet seen a WRITE from the leader acknowledge this message and will ignore the previous leader's message. Those who have will forward this block to the new leader.
- New leader starts its round with an AGREE message as before, but they also include 2f+1 signed acks for their corresponding FREEZE. If any of these acks include a WRITE block then the new leader must propose that block. Otherwise, it proposes its own.

## Correctness Sketch

To prove a BAR-tolerant algorithm correct, the first step is to show that no individual rational actor can improve their utility by deviating from the protocol, under the assumption that the other rational actors are following the protocol.

There are several things the rational actor may care about. First, there is a coin reward for having one's block added to the chain. Actors can't do much to influence the reward. They can follow the protocol so it's back to their turn in a timely manner and they get the reward (that's good). They can't jump the queue - we enforce a strict round-robin rule: everyone gets their turn. They can try to prevent other actors from joining - this is a topic on its own and we discuss it elsewhere. Second, preventing blocks from being added is something a rational actor might cheat for. Sending fake FREEZE messages doesn't work because it requires signatures from other actors. The actor could try sending fake TIMEOUT messages to speed up eviction of a leader that is being slow (or perhaps in a period of asynchrony). This also requires others to cooperate but the argument is more subtle because some non-malicious actors may genuinely observe a leader being slow. We use the clock rate assumption here: the next leader will reject TIMEOUT messages if they arrive too early. Since the clock ratio is r, a message that arrives before T/r is too early. If a rational actor sends at time T then it's guaranteed that the new leader will accept it. In fact this is the earliest time the new leader is guaranteed to accept it, so it's the optimal time to send (hence the algorithm is at equilibrium). Conversely, one might imagine rational actors trying to prevent others from proposing a block. However, actors take turns and they cannot end this round without suspect messages from other actors.

Once we've shown the algorithm is a Byzantine Nash Equilibrium, for the rest of the argument we can safely assume that rational actors will follow the protocol. The key point of the freezing algorithm is that it will not result in two different blocks being decided because (a) if a block could be decided by the previous round, then that block is discovered by the freezing algorithm and proposed in the next round. (b) otherwise, the previous round will never decide a block. The first point follows from the fact that the DECIDE message need 2f+1 signatures: at least one of the signatories will be uncovered by the freezing algorithm and the block will be forwarded to the next round. The second point follows from the symmetric argument that FREEZE is sent to 2f+1 actors. If all accept it before seeing a WRITE, then there are not enough left to sign the DECIDE message so the previous round will not decide.

This algorithm allows the system to pick a block and prevents malicious or rational actors from preventing consensus or interfering with the turn order.

# Conclusion

In this paper we argue for algorithms that tolerate both rational (self-interested) actors, and malicious (Byzantine) ones. These algorithms are called BAR-tolerant. When writing one, follow these three steps: clearly define the utility function for the rational actors, prove the algorithm is such that there is no benefit from unilaterally deviating (that is, it's a Byzantine Nash Equilibrium), then prove the algorithm correct assuming the rational actors follow the protocol. We explain the gatekeeper attack, where members of a system selfishly decide to prevent newcomers from joining. We sketch a BAR-tolerant blockchain protocol. It relies on a strict order to decide who gets to propose a new block (so there's no need to race to solve a crypto puzzle) and it relies on hardware ID tokens to make sure every computer is only represented at most once as a block proposer. It also defends against the gatekeeper attack. The BAR-tolerant approach is naturally also applicable to other blockchain algorithms.